\documentclass{aa}  
\usepackage{graphicx}
\usepackage{natbib}
\bibpunct{(}{)}{;}{a}{}{,} 
\usepackage{txfonts}
\usepackage{hyperref}
\hypersetup{colorlinks=true,linkcolor=blue,citecolor=blue,filecolor=blue,linkcolor=blue}

\begin{document} 

   \title{European VLBI Network observations of the peculiar radio source 4C~35.06 overlapping with a compact group of nine galaxies}
    \titlerunning{EVN observations of 4C 35.06}
   \subtitle{}

   \author{P. M. Veres
          \inst{\ref{rub}, \ref{konkoly}, \ref{mta}}
          \and K. \'E. Gab\'anyi
          \inst{\ref{elte_astro}, \ref{extr_group}, \ref{konkoly}, \ref{mta}}
          \and S. Frey
          \inst{\ref{konkoly}, \ref{mta},  \ref{elte_phys}}
          \and Z. Paragi
          \inst{\ref{jive}}
          \and T. An
          \inst{\ref{shao}, \ref{key}}
          \and J. Bagchi
          \inst{\ref{christ}}
          \and \'A. Bogd\'an
          \inst{\ref{chandra}}
          \and K. G. Biju
          \inst{\ref{wmo}}
          \and E. Kun
          \inst{\ref{rub},\ref{konkoly}, \ref{mta}, \ref{theo},\ref{astropart}}
          \and J. Jacob
          \inst{\ref{newman}}
          \and B. Adebahr
          \inst{\ref{rub}}
          }

   \institute{Astronomisches Institut, Ruhr-Universit\"at Bochum,
Universit\"atsstr. 150, 44801 Bochum, Germany \email{veres@astro.ruhr-uni-bochum.de} \label{rub} \and Department of Astronomy, Institute of Physics and Astronomy, ELTE E\"otv\"os Lor\'and University, P\'azm\'any P\'eter s\'et\'any 1/A, H-1117 Budapest, Hungary \label{elte_astro} \and HUN-REN--ELTE Extragalactic Astrophysics Research Group, ELTE E\"otv\"os Lor\'and University, P\'azm\'any P\'eter s\'et\'any 1/A, H-1117 Budapest, Hungary \label{extr_group} \and Konkoly Observatory, HUN-REN Research Centre for Astronomy and Earth Sciences, Konkoly Thege Mikl\'os \'ut 15-17, H-1121 Budapest, Hungary \label{konkoly} \and CSFK, MTA Centre of Excellence, Konkoly Thege Miklós \'ut 15-17, H-1121 Budapest, Hungary \label{mta} \and Institute of Physics and Astronomy, ELTE E\"otv\"os Lor\'and University, P\'azm\'any P\'eter s\'et\'any 1/A, H-1117 Budapest, Hungary \label{elte_phys} \and Joint Institute for VLBI ERIC, Oude Hoogeveensedijk 4, 7991 PD Dwingeloo, The Netherlands \label{jive} \and Shanghai Astronomical Observatory, Chinese Academy of Sciences, Shanghai 200030, PR China \label{shao} \and Key Laboratory of Radio Astronomy and Technology, Chinese Academy of Sciences, A20 Datun Road, Chaoyang District, Beijing 100101, PR China \label{key} \and Department of Physics and Electronics, CHRIST (Deemed to be University), Hosur Road, Bangalore 560029, India \label{christ} \and 
Center for Astrophysics, Harvard \& Smithsonian, 60 Garden Street, Cambridge, MA 02138, USA \label{chandra} \and W.M.O. Arts \& Science College, Muttil, Wayanad, Kerala 673122, India \label{wmo} \and
Theoretical Physics IV, Faculty for Physics \& Astronomy, Ruhr University Bochum, Universit\"atsstr. 150, 44780 Bochum, Germany\label{theo} \and Ruhr Astroparticle And Plasma Physics Center, Ruhr-Universit\"at Bochum, Universit\"atsstr. 150, 44780 Bochum, Germany \label{astropart} \and Newman College, Thodupuzha, Kerala 685584, India \label{newman}}    
        

   \date{}

 
  \abstract
   {According to the hierarchical structure formation model, brightest cluster galaxies (BCGs) evolve into the most luminous and massive galaxies in the Universe through multiple merger events. The peculiar radio source 4C~35.06 is located at the core of the galaxy cluster Abell 407, overlapping with a compact group of nine galaxies. Low-frequency radio observations have revealed a helical, steep-spectrum, kiloparsec-scale jet structure and inner lobes with less steep spectra, compatible with a recurring active galactic nucleus (AGN) activity scenario. However, the host galaxy of the AGN responsible for the detected radio emission remained unclear.}
   {We aim to identify the host of 4C~35.06 by studying the object at high angular resolution and thereby confirm the recurrent AGN activity scenario.}
   {To reveal the host of the radio source, we carried out very long baseline interferometry (VLBI) observations with the European VLBI Network of the nine galaxies in the group at 1.7 and 4.9~GHz.}
   {We detected compact radio emission from an AGN located between the two inner lobes at both observing frequencies. In addition, we detected another galaxy at 1.7~GHz, whose position appears more consistent with the principal jet axis and is located closer to the low-frequency radio peak of 4C~35.06. The presence of another radio-loud AGN in the nonet sheds new light on the BCG formation and provides an alternative scenario in which not just one but two AGNs are responsible for the complex large-scale radio structure.}
   {}


   \maketitle

\section{Introduction}
\label{sec:intro}

According to the hierarchical structure formation model, galaxies grow via merging. This is especially true for the giant elliptical galaxies (cD galaxies) often residing in the densest, central region of rich galaxy clusters. These brightest cluster galaxies (BCGs) are thought to be formed via the merger events of galaxies devoid of cold interstellar gas, so-called dry mergers \citep[e.g.,][]{bcg_assembly}.

Almost all of the known cD galaxies are radio-loud, and their jets provide feedback in a dense cluster \citep[e.g.,][]{cD_radio}. It is still unknown what triggers the radio emission.\ The presence of multiple lobes in the large-scale radio structure is often explained by invoking recurrent activity \citep[e.g.,][]{recurrent1}, and the complex morphologies (e.g., X-shaped or Z-shaped sources) are interpreted via changes in the central engine  \citep[spin-flip of the supermassive black hole due to a recent merging event; e.g.,][]{spinflip_gergely} and/or interactions with the dense environment around the host galaxy \citep{capetti2002}.

The peculiar radio source 4C~35.06 is located at the centre of the Abell\,407 galaxy cluster, overlapping with a compact group of nine elliptical galaxies that occupy a region with a diameter of $\sim 1\arcmin$. This corresponds to a projected linear size of $\sim 55$\,kpc at the average redshift of the system, $z=0.047$ \citep{cosmocalc}. 

\begin{figure}
    \centering
    \resizebox{\hsize}{!}{
\includegraphics[clip=,bb=100 230 490 620]{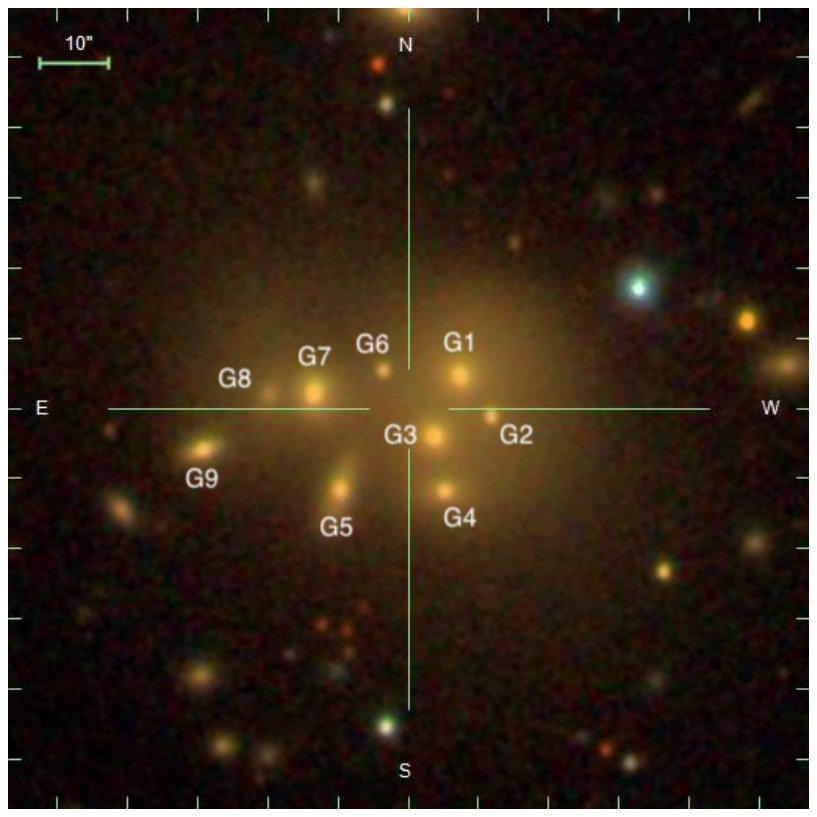}}
    \caption{SDSS optical image of the tight group of nine galaxies and the faint intra-cluster light halo in the centre of the galaxy cluster Abell\,407 \citep{sdss_dr17}. The nine galaxies are labelled G1 to G9, following \cite{biju}.}
    \label{fig:sdss}
\end{figure}

Our target in the Abell\,407 cluster, a compact group of nine galaxies embedded in the vast envelope  ($\sim$100 kpc)
of a faint stellar halo, was first noticed in the 1970s by Fritz Zwicky \citep[V Zw 311;][]{Zwicky}, and \citet[]{biju} later named this extraordinary system ‘Zwicky’s Nonet’ in his honour. To our knowledge, this remarkable grouping is the most compact and richest system of
galaxies (all at similar redshifts) known to date. \cite{Gunn} studied the system in more detail and concluded that this puzzling grouping is an exceptionally rare snapshot of a future cD or a central BCG currently being assembled in situ as a result of tidal disruption and mergers of multiple galaxies. Moreover, the intra-cluster halo of light is composed of stars not bound to any galaxy of the cluster but rather to the gravitational
potential of the cluster as a whole. As the intra-cluster halo of light possibly forms through galaxy encounters in the cluster, it contains vital information about the evolution of clusters and BCG formation \citep{montes}. An optical image with the nine galaxies of Zwicky's Nonet is shown in Fig.~\ref{fig:sdss}. The X-ray properties, optical spectroscopic studies, and the large number of galaxies with similar luminosities and colours suggest an undergoing BCG formation via slow major mergers \citep{formation2,formation,geng_chandra}, making the cluster an excellent laboratory for galaxy evolution studies.

The radio emission of 4C~35.06 was studied with arcsecond-scale resolution by \cite{relic} using Low-Frequency Array (LOFAR) observations and later by \cite{biju} using observations conducted by the Giant Meterwave Radio Telescope (GMRT) and the\textit{ Karl G. Jansky} Very Large Array (VLA). These studies revealed helically twisted jets and outer diffuse lobes extending over $\sim400$ kpc, with steep radio spectra indicating a spectral age of a few times $10^7-10^8$\,yr. The shape of the jet and the concentration of nine galactic nuclei in the central region of the low-frequency radio source suggest a precession of the active galactic nucleus (AGN) jet and gravitational perturbation caused by the nearby galaxies. \cite{biju} hypothesised that G6 may host the jet-emitting radio-loud AGN. However, previous milliarcsecond (mas) resolution Very Long Baseline Array observations by \cite{Liuzzo} taken at $8$\,GHz revealed compact radio emission at the position of G3. We carried out high-resolution very long baseline interferometry (VLBI) observations of the system to look for compact radio emission at the central optical positions of each of the nine galaxies. 

Throughout the paper, we assume a flat $\Lambda$ cold dark matter cosmological model with parameters $H_0=70\,\textrm{km\,s}^{-1}\textrm{\,Mpc}^{-1}$, $\Omega_\Lambda=0.73$, and $\Omega_\textrm{m}=0.27$. The paper is organised as follows. We describe our observations and data analysis in Sect.~\ref{sec:obs}, present the results in Sect.~\ref{sec:results}, and discuss our findings in Sect.~\ref{sec:discussion}. A summary is given in Sect.~\ref{sec:summary}.

\section{Observations and data analysis}
\label{sec:obs}

Our dual-frequency phase-referenced European VLBI Network (EVN) observations of the peculiar radio source 4C~35.06 were carried out at $1.7$\,GHz on 2017 May 29 and at $4.9$\,GHz on 2017 June 12 (project code: EG098, PI: K. Gabányi). Each of the eight intermediate-frequency channels was divided into 32 (at $1.7$~GHz) and 64 (at $4.9$~GHz) $500$\,kHz spectral channels, resulting in total bandwidths of $128$ and $256$\,MHz, respectively. The data were recorded in left- and right-hand circular polarisations with data rates of $1$ and $2$~Gbit\,s$^{-1}$ at $1.7$ and $4.9$\,GHz, respectively. To potentially image the compact radio emission of the nine galaxies that are located within the primary beam of the radio telescopes, we used a single pointing for all targets and requested multiple-phase-centre correlation \citep[e.g.][]{multi-p-centre-morgan,multi-p-centre-cao}. The pointing coordinates were those of G3, taken from \citet{Liuzzo} at right ascension $03^\mathrm{h} 01^\mathrm{m} 51\fs813$ and declination $+35\degr 50\arcmin 19\farcs587$. The Sloan Digital Sky Survey (SDSS) optical coordinates were taken as a priori values for the other galaxies for the correlation, performed by the EVN Data Processor \citep{JIVEcorrelator} at the Joint Institute for VLBI European Research Infrastructure Consortium (Dwingeloo, The Netherlands) with $2$\,s averaging time. The total observing times of $10$\,h and $5$\,h resulted in on-source times of $\sim 7$\,h and $\sim 3.5$\,h at $1.7$ and $4.9$\,GHz, respectively. The participating antennas are listed in Table~\ref{table:antennas}. The on-target intervals were interleaved with observations of two calibrator sources. The nearby quasar J0304$+$3348, at an angular distance of $2\fdg11$, was observed as a phase-reference calibrator, and the bright quasar J0230$+$4032 was observed for fringe finding. Archival VLBI maps\footnote{\url{http://astrogeo.org/cgi-bin/imdb_get_source.csh?source=J0304\%2B3348}} of J0304$+$3348 indicate a prominent core--jet structure, with the jet pointing towards the south-east.

\begin{table}
\caption{Participating antennas for the EVN observations.} 
\centering
\label{table:antennas}      
\centering                                      
\begin{tabular}{c c}          
\hline\hline                        
Observing frequency & Participating antennas \\    
(GHz) & \\
\hline                                   
    $1.7$ & O8, Ur, Tr, Sv, Zc, Bd, \\
    experiment EG098A & Ir, Jb, Mc, Ef, Wb, Hh\\
    \hline 
    $4.9$ & Jb, Ef, Mc, O8, T6, Ur, Tr, Sv,\\
    experiment EG098B &  Zc, Bd, Ir, Nt, Ys, Wb, Hh \\
\hline                                             
\end{tabular}
\tablefoot{Antennas are: Badary, Russia (Bd), Effelsberg, Germany (Ef), Hartebeesthoek, South Africa (Hh), Irebene, Latvia (Ir), Jodrell Bank Mk2, United Kingdom (Jb), Noto, Italy (Nt), Onsala, Sweden (O8), Svetloe, Russia (Sv), Tianma, China (T6), Toru\'n, Poland (Tr), Urumqi, China (Ur), Westerbork, the Netherlands (Wb), Yebes, Spain (Ys), and Zelenchukskaya, Russia (Zc).}
\end{table}

We calibrated the data using the U.S. National Radio
Astronomy Observatory Astronomical Image Processing System \citep[\textsc{AIPS};][]{aips} software package. During the data analysis, we followed the standard procedures when first calibrating the dataset that contains the data of one of the phase centres at the location of G1 and the calibrator sources. These steps were the amplitude calibration using the measured system temperatures\footnote{Since the station Urumqi (Ur) had no system temperature information available for experiment EG098A, we used nominal values for amplitude calibration.} and gain curves of each telescope, corrections for the dispersive ionospheric delays and for the parallactic angle, and global fringe-fitting for the calibrator sources. We imaged the calibrator sources with the \textsc{Difmap} software \citep{difmap}, using the hybrid mapping technique with \textsc{clean} deconvolution \citep{hybridmapping}. After several iterations of \textsc{clean} and phase self-calibration, we performed an overall amplitude self-calibration, obtaining antenna-specific gain correction factors. We then applied amplitude corrections for telescopes with $>5\%$ deviation from the gain solutions in \textsc{AIPS}. Using the \textsc{clean} component model we built up in \textsc{Difmap}, we repeated the fringe-fitting in \textsc{AIPS} for the phase-reference calibrator source to correct for the source structure since it contributes to the measured interferometric phases. Finally, we interpolated the solutions of the last fringe-fitting to the data of the target source located at the phase centre of G1, and exported the calibrated visibility file from \textsc{AIPS} to \textsc{Difmap} for later imaging.

Then we calibrated the visibilities associated with the other eight phase centres using \textsc{AIPS}. First, we copied the necessary calibration tables containing the amplitude calibrations to each of the other visibility files supplied by the correlator. Then we performed corrections for the dispersive ionospheric delays and for the parallactic angle. As the next step, we copied the last phase solutions obtained earlier for the phase-reference calibrator source. This calibration table also contained the antenna gain corrections. We applied the solutions to each remaining target source. Finally, we exported the calibrated visibility data of the additional eight target sources for imaging in \textsc{Difmap}. 

Imaging at the nine phase centres was performed in \textsc{Difmap}. Due to the low peak intensity values in the maps, we did not attempt any self-calibration. For the detected targets, after flagging outlier data points, we fitted circular Gaussian model components \citep{modelfit} to the visibility data to describe the brightness distribution of the source. Finally, to decrease the noise level and consequently improve the signal-to-noise ratio (S/N), we performed $1000$ \textsc{clean} iterations in the residual images with a small loop gain of $0.01$.

\section{Results}
\label{sec:results}

Out of the nine positions, we detected significant radio emission, defined as exceeding $6$ times the local noise level ($\sigma$), at the phase centres of G1 at $1.7$\,GHz only, and of G3 at both frequencies. For the other seven phase centres, we provide $6\sigma$ brightness upper limits. When using empty-field EVN images, we found that there might occasionally be spurious features at or slightly above the $5\sigma$ rms noise level; therefore, we considered $6\sigma$  a safer threshold for detection. We note that a $6\sigma$ rms noise level is generally used as the detection threshold for wide-field EVN surveys \citep[e.g.][]{det1,det2}. The upper limits for the seven other galaxies are $\sim 0.1$\,mJy\,beam$^{-1}$ and $\sim 0.06$\,mJy\,beam$^{-1}$ at $1.7$\,GHz and $4.9$\,GHz, respectively. Therefore, there are no compact unresolved radio sources at the centres of these galaxies with flux densities $\ga 100\,\mu$Jy at $1.7$\,GHz and $\ga 60\,\mu$Jy at $4.9$\,GHz.

\subsection{Radio emission in G1}
\label{subsec:results_g1}
We detected a single feature with a S/N of $6.3 \sigma$ in the $1.7$ GHz image of G1. The right ascension and declination of the radio feature are $\alpha_\textrm{G1}= 03^\mathrm{h} 01^\mathrm{m} 51\fs5127 \pm 0\fs0003$ and $\delta_\textrm{G1}= +35\degr 50' 28\farcs294\pm 0\farcs004$, respectively. When calculating the VLBI positional uncertainty, we took into account the positional uncertainty of the phase calibrator source, its angular separation from the target source, and the resolution of the observations, which depends on the thermal noise of the interferometric images \citep{arrayerror}. The position of the phase calibrator source is listed with an error of $0.12$\,mas in the third realisation of the International Celestial Reference Frame (ICRF3; \citealt{icrf3}). Following \cite{arrayerror}, the thermal noise-limited positional accuracy is $0.29$\,mas. The angular separation between the target and the calibrator source implies a positional error of about $4$\,mas at $1.7$\,GHz \citep{Chatterjee}. The estimated positional uncertainty of $4$\,mas is therefore dominated by the uncertainty arising from the angular separation between the calibrator and the target sources. 

\begin{figure}
  \resizebox{\hsize}{!}{\includegraphics{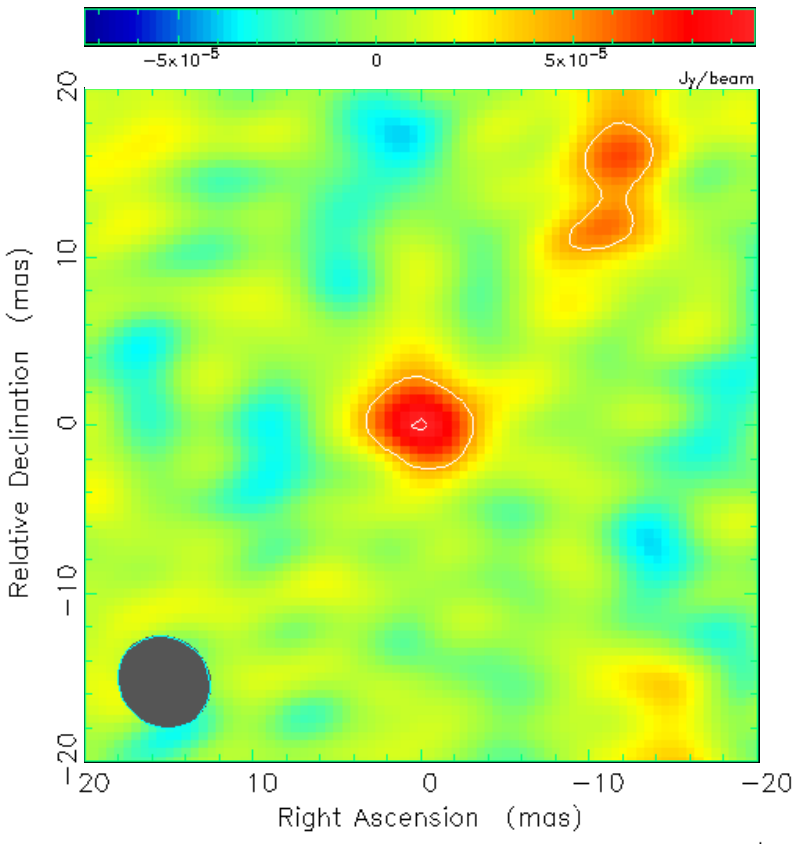}}
  \caption{Naturally weighted $1.7$ GHz EVN map of galaxy G1. The grey ellipse seen in the lower-left corner represents the size of the Gaussian restoring beam, $5.72\textrm{\,mas} \times 5.24$\,mas (FWHM), at a major axis position angle of PA $=46\fdg5$. The contour levels are drawn at around $3\sigma$ (corresponding to $0.05 \textrm{\,mJy\,beam}^{-1}$) and $6\sigma$ image noise levels. The peak brightness is $0.1\textrm{\,mJy\,beam}^{-1}$.}
  \label{pic:G1}
\end{figure}

The final naturally weighted image of G1 is displayed in Fig.~\ref{pic:G1}. Our VLBI map shows an unresolved source without any prominent jet structure. The radio emission can be well described by a single circular Gaussian brightness distribution model. As phase self-calibration was not possible, due to the imperfect phase calibration (i.e. switching between the target and reference sources), a coherence loss \citep{coherence-loss} decreases the derived flux density value of the target. Based on previous EVN observations of faint targets \citep[e.g.][]{mosoni,triple}, we estimate the flux density reduction caused by the coherence loss to be $\sim 25$\,\%. The parameters of the component  given in Table \ref{table:modelfit} account for this coherence loss. We used the relations of \cite{Fomalont} to calculate the uncertainties of the flux density and size parameters of the fitted component.

\subsection{Radio emission in G3}
\label{subsec:results_g3}
We detected radio emission in G3 with S/Ns of $\sim27 \sigma$ and $\sim45 \sigma$ at $1.7$\,GHz and $4.9$\,GHz, respectively. The $4.9$ GHz right ascension and declination coordinates of the detected feature are $\alpha_\textrm{G3}= 03^\mathrm{h} 01^\mathrm{m} 51\fs8129 \pm 0\fs0001$ and $\delta_\textrm{G3}= +35\degr 50' 19\farcs588 \pm 0\farcs002$, respectively. We calculated the positional uncertainties in a similar way as described in Sect.~\ref{subsec:results_g1}, and give the more accurate 4.9 GHz coordinates here. The thermal noise-limited position determination error ($8$\,$\mu$as), the positional uncertainty of the calibrator source ($0.12$\,mas), and the uncertainty originating from the angular separation between the calibrator and the target source at $4.9$\,GHz ($1.5$\,mas) result in an overall positional error of $1.5$\,mas. The coordinates agree with those derived from the $1.7$ GHz EVN data within the uncertainties.

\begin{figure}
  \resizebox{\hsize}{!}{\includegraphics{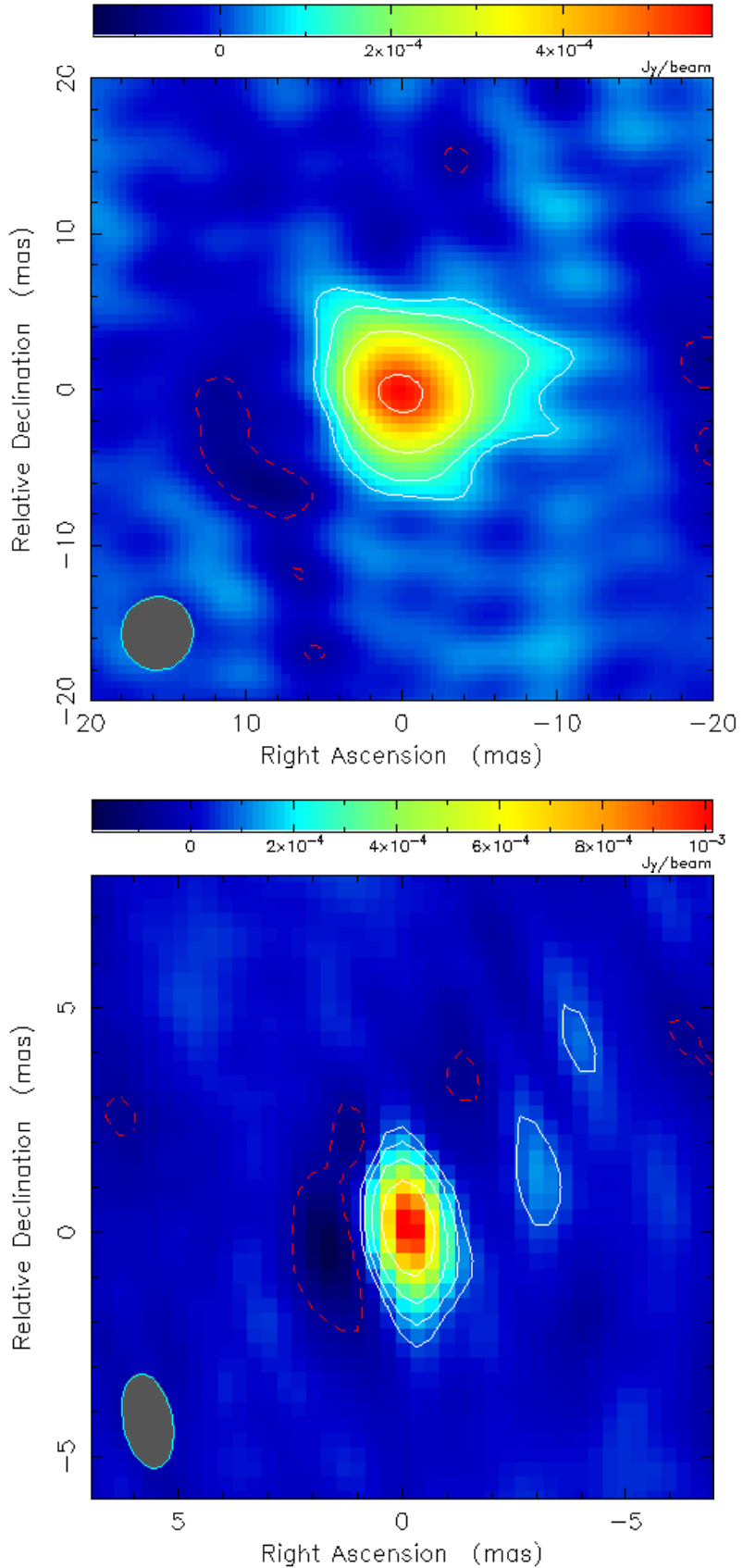}}
  \caption{Naturally weighted $1.7$ GHz \textit{(top)} and $4.9$ GHz \textit{(bottom)} EVN maps of galaxy G3. The grey ellipses in the lower-left corners represent the Gaussian restoring beams. Their parameters are $4.75\textrm{\,mas} \times 4.61$\,mas (FWHM) at a major axis position angle of PA $=-16\fdg4$ at $1.7$\,GHz, and $2.13\textrm{\,mas} \times 1.10$\,mas (FWHM) at a major axis position angle of PA $=11\fdg1$ at $4.9$\,GHz. The lowest contour levels are drawn at $\pm 3\sigma$ image noise levels. They are $0.06\textrm{\,mJy\,beam}^{-1}$ at $1.7$\,GHz and $0.07\textrm{\,mJy\,beam}^{-1}$ at $4.9$\,GHz. Further positive contour levels increase by a factor of $2$. The peak brightnesses are $0.6\textrm{\,mJy\,beam}^{-1}$ at $1.7$\,GHz and $1.0\textrm{\,mJy\,beam}^{-1}$ at $4.9$\,GHz.}
  \label{pic:G3}
\end{figure}

We present the final naturally weighted images of G3 in Fig.~\ref{pic:G3}. Our EVN map taken at $1.7$\,GHz shows an elongated feature with an extension to the west. However, no prominent jet structure can be seen. At $4.9$\,GHz, the radio emission is almost completely point-like, with a hint of an extension towards the west.

At both frequencies, a single circular Gaussian component fitted to the visibility data can adequately describe the brightness distribution. The parameters of the fitted components, with the flux densities increased by $25$\% to account for the coherence loss, are given in Table~\ref{table:modelfit}. Their uncertainties were calculated using the relations from \cite{Fomalont}.

\begin{table}
\caption{Parameters of the best-fit Gaussian model components of the detected radio features. }              
\label{table:modelfit}      
\centering                                      
\begin{tabular}{c c c c}          
\hline\hline                        
ID & Frequency & Flux density & Size \\    
 & (GHz) & (mJy) & (mas) \\
\hline                                   
    G1 & 1.7 & $0.14 \pm 0.05$ & $1.9 \pm 0.5$ \\
    \hline
    G3 & 1.7 & $2.0 \pm 0.5$ & $6.2 \pm 1.7$ \\
       & 4.9 & $1.4 \pm 0.4$ & $0.5 \pm 0.1$ \\
\hline                                             
\end{tabular}
\tablefoot{The fitted flux densities have been increased by $25$\,\% to account for the coherence loss. `Size' is the FWHM diameter of the fitted component.}
\end{table}

\section{Discussion}
\label{sec:discussion}

The coordinates of the radio features detected in G1 and G3 agree within their uncertainties with the {\it Gaia} \citep{gaia} Data Release 3 \citep[DR3;][]{gaiadr3} position of the respective galaxies. Thus, we have identified the radio sources with the optical galaxies. We note that, in addition to the standard positional error, the overall {\it Gaia} DR3 positional uncertainty is strongly influenced by the astrometric excess noise parameter\footnote{\url{https://gea.esac.esa.int/archive/documentation/GDR2/Gaia_archive/chap_datamodel/sec_dm_main_tables/ssec_dm_gaia_source.html}} for both galaxies. This parameter characterises the goodness of the fit of the best-fitting standard astrometric model to the observations \citep{excess}. The relatively high value of this parameter is a potential signature of multiple systems, for example X-ray binaries \citep{excess2} and exoplanet systems. In our case, the high astrometric excess noise values, $50$\,mas for G1 and $33$\,mas for G3, are fully consistent with the multiple nature of the cluster.

\subsection{Brightness temperatures of G1 and G3}
\label{subsec:Tb}
We calculated the brightness temperature of the radio-emitting feature in G1 at $1.7$\,GHz and G3 at $4.9$\,GHz using the following equation \citep{Tb, Tb2}: 
\begin{equation}
T_{\mathrm{b}} = 1.22 \times 10^{12} (1+z) \frac{S}{\theta^{2} \nu^{2}}\,\textrm{K,}
\end{equation}
where $S$ is the flux density of the fitted Gaussian component measured in Jy, $\theta$ is the full width at half maximum (FWHM) diameter of the fitted component in mas, and $\nu$ is the observing frequency in GHz. The obtained brightness temperatures are 
$T_{\mathrm{b}}^\mathrm{G1} = (1.7 \pm 1.0) \times 10^{7}$\,K for G1 at $1.7$\,GHz and $T_{\mathrm{b}}^\mathrm{G3} = (3.0 \pm 1.5) \times 10^{8}$\,K for G3 at $4.9$\,GHz. These values significantly exceed $\sim 10^5$\,K, known as the upper limit for star-forming galaxies \citep{sfrTb}, indicating that the radio emission originates from AGN activity in both sources.

\subsection{Spectral index calculations for G1 and G3}
\label{subsec:spx}
We obtained a spectral index of $\alpha_\textrm{G3} = -0.34 \pm 0.36 $ (where $\alpha$ is defined as $S \propto \nu^{\alpha}$) for galaxy G3 between $1.7$\,GHz and $4.9$\,GHz using the flux densities of the fitted Gaussian model components. The relatively flat radio spectrum suggests that the radio emission most likely originates from the compact jet base. Since we detected G1 only at $1.7$\,GHz but not at $4.9$\,GHz, we give an upper limit of $\alpha_\textrm{G1} < -0.8$ for the spectral index of the radio emission feature.

\subsection{Radio-emitting AGNs in Abell 407}
Based on multi-frequency radio observations, galaxies G3 \citep{relic,Liuzzo} and G6 \citep{biju} were favoured as potential hosts of the radio-loud AGN responsible for the large-scale radio structure and as such are associated with the peculiar radio source 4C~35.06. Our detections and the estimated brightness temperature values of G1 and G3 suggest that two radio-emitting AGNs are located at the centre of the galaxy cluster.

\begin{figure}
  \resizebox{\hsize}{!}{\includegraphics{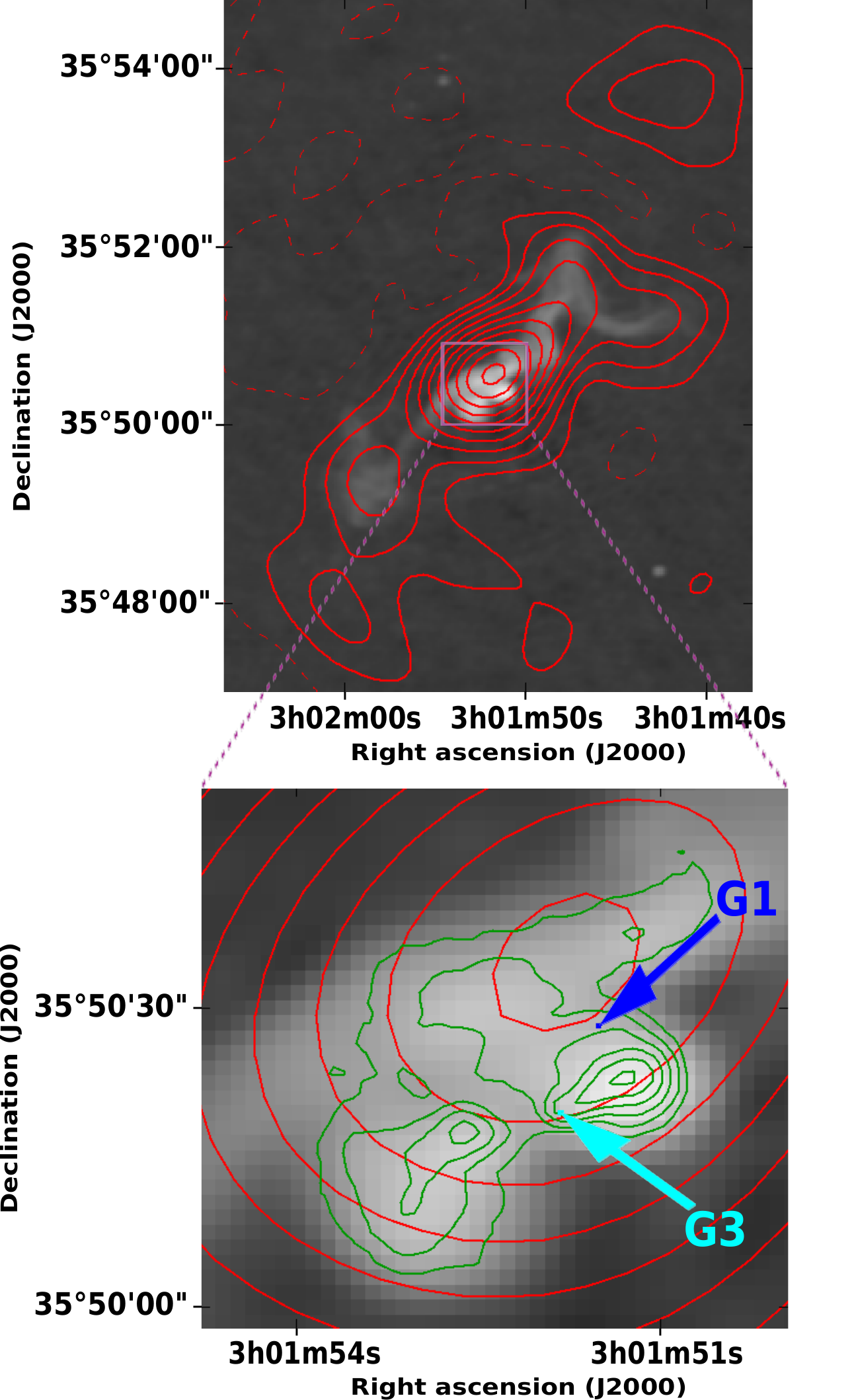}}
  \caption{Overlay maps of archival radio data of the radio source 4C~35.06, indicating the positions of the two radio AGNs detected with the EVN. \textit{Upper panel:} $610$\,MHz GMRT observations published by \cite{biju}, shown with the grayscale map, and  the $62$\,MHz observations published by \cite{relic}, shown with the red contours (ten levels spaced over a square root interval from $-5\sigma$ to $183\sigma$)  .  \textit{Bottom panel:} Zoomed-in view of the central region of 4C~35.06. The green contours represent the most recent VLASS observations (VLASS2.2 observing campaign),  confirming the presence of the inner lobes. The peak brightness of this map is $14.3$\,mJy\,beam$^{-1}$, while the noise level is $0.14$\,mJy\,beam$^{-1}$. The seven contour levels span the range $-3\sigma$ to $100\sigma$. The dark blue and light blue arrows indicate the EVN positions of galaxies G1 and G3, respectively.}
  \label{pic:overlay}
\end{figure}

Galaxy G3 is located between two arcsecond-scale radio features observed with the GMRT at $610$\,MHz and with the VLA within the Very Large Array Sky Survey \cite[VLASS;][]{vlass} at $3$\,GHz as radio intensity peaks near the optical positions of galaxies G2 and G5 (see Figs.~\ref{fig:sdss} and \ref{pic:overlay}). The radio features can also be seen in the $5$ GHz VLA maps published by \cite{relic}, who identified this bipolar structure as inner radio lobes. Our EVN maps of G3 (Fig.~\ref{pic:G3}) suggest an elongated feature with an extension to the west, consistent with the findings of \cite{Liuzzo} and with the orientation of the inner lobes. The jet activity of G3 provides a plausible explanation for the origin of these inner lobes; however, the galaxy is offset by $\sim10\arcsec$  to the south from the large-scale jet axis (Fig.~\ref{pic:overlay}).

\cite{relic} interpreted the puzzling structure of 4C~35.06, the inner twin lobes with a less steep spectrum and the outer diffuse lobes, as being due to recurrent AGN activity (i.e. two subsequent episodes of jet activity). In this scenario, galaxy G3 moved some distance to the position where we observe it now, potentially explaining the orientation offset seen between the inner lobes and the outer regions, as well as the large-scale helical jet pattern.

Our detection of a second radio-emitting AGN in the nonet offers an alternative explanation in which the relatively young inner lobes and the more extended, old, steep-spectrum emission are not associated with the same galaxy. Galaxy G1 is located to the north-west of G3, and its position in the cluster is more consistent with the principal axis of the large-scale jet (Fig.~\ref{pic:overlay}). Additionally, the LOFAR $62$ MHz radio peak (see Fig.~\ref{pic:overlay}) is seen closer to G1 than to G3. Archival Westerbork Synthesis Radio Telescope (WSRT) $610$ MHz observations of the galaxy cluster Abell 407 also indicate a low-frequency radio peak located at the position of the north-westernmost galaxy, G1 \citep{wsrt}. These findings suggest that the steep-spectrum relic lobes are connected to a past activity episode of G1. This is consistent with our EVN image of the galaxy (Fig.~\ref{pic:G1}) showing no prominent jet structure and a low level of activity. In summary, in this scenario, the double--double radio morphology is the outcome of the jet activities of both G1 and G3. Notably, these two galaxies are the two optically brightest members of the nonet and, together with G5, have the highest supermassive black hole masses ($M^{\mathrm{G1}}_{\mathrm{SMBH}} \sim4 \times 10^{8} M_{\odot}$ and $M^{\mathrm{G3}}_{\mathrm{SMBH}} \sim10^{9} M_{\odot}$;  \citealt{biju}), providing further support for their radio AGN nature.

The \textit{Chandra} Advanced CCD Imaging Spectrometer (ACIS) detected both G1 and G3, together with five other galaxies of the system, in the $0.5-2$\,keV and $2-7$\,keV X-ray bands \citep{geng_chandra}. \cite{geng_chandra} also revealed that galaxy G3 has the most prominent X-ray thermal corona.

We note that multiple-AGN activity is common in galaxy clusters \citep{2020MNRAS.498.2719T}. The destabilised gas inflow towards the inner regions supplies the central black hole with material to be accreted, resulting in a higher accretion rate \citep{2018Natur.563..214K}. The projected distance between G1 and G3 is $\sim 9$\,kpc. At this distance, the slow merging process and the AGN activity play a crucial role in the galaxy evolution \citep{2020MNRAS.498.2719T}. However, multiple-AGN activity in galaxy clusters is significantly more common in high-redshift galaxy cluster cores than in low-redshift clusters like Abell 407 \citep{2013ApJ...768....1M}.

Regardless of the contribution of G1 to the overall radio emission of 4C~35.06, the ongoing formation of the cD galaxy provides a reasonable explanation for the helical jet pattern found by \cite{biju}.

\section{Summary}
\label{sec:summary}

Zwicky's Nonet is a compact group of nine galaxies within $\sim1\arcmin$ of one another in the central region of the galaxy cluster Abell 407. It is associated with the extended radio source 4C~35.06 and shows a complex jet structure. 

We observed all nine galaxies in this group at high angular resolution with the EVN to look for the AGN that is responsible for the radio emission. With the mas-scale resolution of the EVN, we detected compact radio-emitting features in two galaxies, G1 and G3. Both have brightness temperatures exceeding the value expected from star-forming galaxies, which must be due to nuclear activity. The brighter radio source of the two, located in G3, was detected at both $1.7$ and $4.9$\,GHz and shows a flat radio spectrum. The weaker one in G1 has a steeper radio spectrum and remains undetected at the higher frequency of $4.9$\,GHz. 

An alternative to the originally proposed scenario to explain the complex large-scale radio structure of 4C~35.06 (i.e. two phases of recurrent activity in the G3 nucleus) is as follows: the extended radio structure could arise from jet emission from two radio-emitting active nuclei in the galaxies of the compact group, as suggested by the VLBI detection of G1. Sensitive radio imaging observations at intermediate angular resolution, for example with the enhanced Multi-Element Remotely Linked Interferometer Network (e-MERLIN), could further reveal details of the jet structure at $\sim 0\farcs1$ scales and thus allow us to confirm one of these two scenarios.

\section*{Acknowledgements}
We gratefully thank the Referee and the Language Editor for their comments and suggestions.

We thank Aleksandar Shulevski (ASTRON, the Netherlands) for sharing the LOFAR $62$-MHz radio map with us.

The European VLBI Network is a joint facility of independent European, African, Asian, and North American radio astronomy institutes. Scientific results from data presented in this publication are derived from the following EVN project code: EG098. The research leading to these results has received funding from the European Commission Horizon 2020 Research and Innovation Programme under grant agreement No. 730562 (RadioNet).

PMV and BA acknowledge support from the German Science Foundation DFG, via the Collaborative Research Center SFB1491: Cosmic Interacting Matters - from Source to Signal. EK thanks the Alexander von Humboldt Foundation for its Postdoctoral Fellowship. KGB and JJ acknowledge support under Associateship Program of
Inter-University Center for Astronomy and Astrophysics (IUCAA), Pune. JB is supported by CHRIST (Deemed to be University), Bangalore. \'AB acknowledges support from the Smithsonian Institution and the Chandra X-ray Center through NASA contract NAS8-03060.

This work was supported by the Hungarian National Research, Development and Innovation Office (NKFIH, grant number OTKA K134213). PMV is grateful for the support received from the observatory assistant programme of the Konkoly Observatory \citep{konkoly}. This project has received funding from the HUN-REN Hungarian Research Network.

Funding for the Sloan Digital Sky 
Survey IV has been provided by the 
Alfred P. Sloan Foundation, the U.S. 
Department of Energy Office of 
Science, and the Participating 
Institutions. SDSS-IV acknowledges support and 
resources from the Center for High 
Performance Computing  at the 
University of Utah. The SDSS 
website is www.sdss4.org. SDSS-IV is managed by the 
Astrophysical Research Consortium 
for the Participating Institutions 
of the SDSS Collaboration including 
the Brazilian Participation Group, 
the Carnegie Institution for Science, 
Carnegie Mellon University, Center for 
Astrophysics | Harvard \& 
Smithsonian, the Chilean Participation 
Group, the French Participation Group, 
Instituto de Astrof\'isica de 
Canarias, The Johns Hopkins 
University, Kavli Institute for the 
Physics and Mathematics of the 
Universe (IPMU) / University of 
Tokyo, the Korean Participation Group, 
Lawrence Berkeley National Laboratory, 
Leibniz Institut f\"ur Astrophysik 
Potsdam (AIP),  Max-Planck-Institut 
f\"ur Astronomie (MPIA Heidelberg), 
Max-Planck-Institut f\"ur 
Astrophysik (MPA Garching), 
Max-Planck-Institut f\"ur 
Extraterrestrische Physik (MPE), 
National Astronomical Observatories of 
China, New Mexico State University, 
New York University, University of 
Notre Dame, Observat\'ario 
Nacional / MCTI, The Ohio State 
University, Pennsylvania State 
University, Shanghai 
Astronomical Observatory, United 
Kingdom Participation Group, 
Universidad Nacional Aut\'onoma 
de M\'exico, University of Arizona, 
University of Colorado Boulder, 
University of Oxford, University of 
Portsmouth, University of Utah, 
University of Virginia, University 
of Washington, University of 
Wisconsin, Vanderbilt University, 
and Yale University.

\bibliographystyle{aa} 
\bibliography{main} 

\end{document}